\documentclass[10pt,tightenlines,eqsecnum,floats,aps,amsmath,amssymb,showpacs,nofootinbib,superscriptaddress,twocolumn,prd]{revtex4}

\usepackage{graphicx}

\usepackage{euscript,amssymb}
\usepackage{mathrsfs}

\usepackage{amsbsy}
\usepackage{epsfig}
\usepackage{color}

\usepackage{amsfonts}
\usepackage{enumitem}
\usepackage{amsmath}

\usepackage{amssymb}

\usepackage{fancyhdr}

\usepackage{esint}
\usepackage[unicode=true, pdfusetitle,
 bookmarks=true,bookmarksnumbered=false,bookmarksopen=false,
 breaklinks=false,pdfborder={0 0 1},backref=false,colorlinks=false]
 {hyperref}

\begin{document}

\title{Quantum theory of electromagnetic fields  in a cosmological quantum spacetime}

\author{Jerzy Lewandowski}
\email{jerzy.lewandowski@fuw.edu.pl}
\affiliation{Faculty of Physics, University of Warsaw, Pasteura 5, 02-093 Warsaw, Poland}

\author{Mohammad Nouri-Zonoz}
\email{nouri@ut.ac.ir}
\affiliation{Department of Physics, University of Tehran, 14395-547 Tehran, Iran}

\author{Ali  Parvizi}
\email{a.parvizi@ut.ac.ir}
\affiliation{Department of Physics, University of Tehran, 14395-547 Tehran, Iran}

\author{Yaser Tavakoli}
\email{yaser.tavakoli@ut.ac.ir}
\affiliation{Department of Physics, University of Tehran, 14395-547 Tehran, Iran}
\affiliation{School of Engineering Science, College of Engineering, University of Tehran, 11155-4563 Tehran, Iran}
\affiliation{School of Physics, Institute for Research in Fundamental Sciences (IPM), 19395-5531 Tehran, Iran}

\begin{abstract}

The theory of quantum fields propagating on an isotropic cosmological quantum 
spacetime is reexamined by generalizing  the scalar test field to an electromagnetic (EM) vector field. For 
any given polarization of the EM field on the classical background, the Hamiltonian can be written 
in the form of the Hamiltonian of a set of  decoupled harmonic oscillators, each  corresponding
to a single  mode of the field. In transition from the classical to quantum spacetime background, following 
the technical procedure  given by Ashtekar {\em et al.} [Phys. Rev. D 79, 064030 (2009)], a quantum theory of the test EM field on an effective 
(dressed) spacetime emerges. The nature of this emerging dressed geometry is independent of the 
chosen polarization, but it may depend on the energy of the corresponding field   mode. Specifically, when the backreaction of the field on the quantum geometry is negligible (i.e.,  a test field approximation is assumed), all field  modes  probe the same effective background independent of the mode's energy. However, when  the backreaction of the  field  modes on the quantum geometry is significant, by employing a Born-Oppenheimer approximation, it is shown that 
a rainbow  (i.e., a mode-dependent) metric emerges. 
The emergence of this  mode-dependent background in the Planck regime may have 
a significant  effect on the creation of quantum particles. The production amount on the dressed 
background is computed and is compared with the familiar results on the classical geometry.

\end{abstract}

\date{\today}

\pacs{04.60.-m, 04.60.Pp, 98.80.Qc}

\maketitle


\section{Introduction}

It is argued that the theory of quantum fields in the Planck regime requires quantization of  the given background spacetime \cite{Ashtekar:2004,Rovelli:2004,Thiemann:2007}. It is then that one could study the dynamics of  quantum fields by analyzing  the field   modes propagating on these  quantum geometries. 
This issue has already  been studied when a test 
{\em scalar field} propagates  on a quantum isotropic \cite{AKL:2009} 
and the simplest anisotropic 
\cite{DLT:2012} background spacetime. 
An extension of those models was also provided in Ref.~\cite{Andrea:2013a}.

In the loop quantum cosmological background \cite{Ashtekar:2003,Ashtekar:2011},  one regards the background matter source $T$ (i.e., given by a scalar  field  \cite{Ashtekar:2006a} or a dust field \cite{Pawlowski:2012})   as a  global {\em relational} 
time variable with respect to which physical observables evolve.
This  simplifies the task of solving the constraints and constructing the physical sector of the theory and  enables one to introduce convenient Dirac observables for the quantum theory.
Then, on the full  phase space of the loop quantum cosmology (LQC), the background geometry is described by a state $\Psi_o$ that  evolves, with respect to $T$, via a Hamiltonian $\hat{H}_o$.  The state 
$\Psi_o$ undergoes a quantum bounce at some time  $T=T_{\rm B}$.
The quantum state of inhomogeneous fields $\varphi$, propagating on this quantum background, 
is denoted by  $\psi$,  which possesses   a natural initial condition at $T_{\rm B}$.

When the massless scalar  inhomogeneities  are regarded as   perturbations, 
the backreaction between fields and the geometry can be discarded,
and thus their state $\psi$ can be  chosen such that their energy density  is small at $T_{\rm B}$. If this situation continues to sustain  at  later  times, then $\Psi_o\otimes\psi$ would be a self-consistent solution to  the total Hamiltonian constraint  of the gravity-field  system, for all times. 
In this case, each {\em massless} mode of the scalar field on the quantum background  probes an effective dressed  geometry that  is independent of the wave number of the mode. Moreover, the emergent geometry has the same isotropy of the original background metric. For massive modes, however, quantum effects  may give rise to  a  small deviation for  the isotropy of the background \cite{Andrea:2013b}. Furthermore, in some situations, a rainbow metric may emerge, a  metric that  has a dependence on the energy of the field  modes propagating on it \cite{Andrea:2015,Andrea:2016}.
Including  the backreaction, as $\Psi_o\otimes\psi + \delta\Psi$,  can  have a significant effect on the total state of the geometry-field system,  which  may  also lead   to  violation of the local Lorentz symmetry \cite{DLT:2012}.
On the other hand, such a quantum geometry provides only probability amplitudes for various homogeneous metrics, and thus  we no longer have a sharply defined, proper, or conformal time variable \cite{Ashtekar:2006a}. This issue was resolved by  studying the dynamics of inhomogeneous perturbations on a quantum Friedmann-Lema\^{i}tre-Robertson-Walker (FLRW) spacetime,  by deparametrizing the Hamiltonian constraint in the background, homogeneous sector \cite{Agullo:2012}. 

After constructing quantum electrodynamics  on the flat spacetime, it is of interest to generalize the formulation to the general curved spacetimes. Illness of the notion of a  particle and lack of 
Poincar\'{e}  symmetry in general spacetime makes it difficult to study quantum theory of fields in spacetimes without preferred symmetries associated with a killing vector. Cosmological backgrounds are among the first targets  in studying the quantization of electromagnetic (EM) fields in curved backgrounds; any observation  made in cosmological scales is related to the analyses of  the EM fields, from classical to quantum phenomenons. We are especially interested in investigating the behavior of quantum EM fields on cosmological quantum backgrounds. There are many quantum gravitational phenomena  occurring in the cosmos, and EM fields originating  from them carry  valuable information about quantum gravitational effects taking  place in that part of the Universe. There are significant studies proposing the violation of the Lorentz symmetry  due to quantum gravity effects, such as spectral lag in gamma-ray bursts and  observations indicating  that the speed of light in vacuum depends on the energy of the photons  \cite{Amelino-Camelia:2013, Amelino-Camelia:1998,Mattingly:2005} (for further phenomenological aspects of quantum gravity, see Refs.~\cite{Smolin:2006,Lafrance:2011,Magueijo:2008,Lafrance:1995,Magueijo:2004}).  It was  also shown that quantum gravity effects may also imprint signatures on the cosmic microwave background observations \cite{Tsujikawa:2004}.  
All of these phenomena  convince us to  study the quantum theory of EM fields  not only on the classical cosmological spacetime but also on the quantum cosmological backgrounds.

The paper is organized  as follows. In Sec.~\ref{sec-BianchiI-class}, we  study the Hamiltonian formalism of a quantum EM vector field on a flat FLRW background. In Sec.~\ref{QG-QFT}, we will present a quantum background on which the EM perturbations can propagate. We will discuss two particular situations: the case in which the backreaction can be neglected and the case in which the backreaction effect is significant in the quantum gravity regime. We will show that, in the presence of backreaction, a deviation from the local Lorentz symmetry emerges. In Sec.~\ref{Particle-Production}, we will discuss  phenomenological aspects of the Lorentz symmetry breaking on the creation of particles in quantum spacetime and then compare our results  with the well-known studies of quantized fields on a classical isotropic background.  Finally, in Sec.~\ref{Conclusion} comes  the conclusion and discussion of our work.

\section{Quantum theory of radiation  field  in a classical  spacetime}
\label{sec-BianchiI-class}

In this section, we will study the  Hamiltonian formulation of an EM field propagating in a flat FLRW background spacetime.

\subsection{EM field equation in flat FLRW spacetime: Radiation gauge}

We consider  a four-dimensional  curved  background spacetime, which is  equipped with the coordinates $(x_0, \mathbf{x})$, where the spatial coordinates $\mathbf{x}\in (0, \ell)$ on a torus  $\mathbb{T}^3$ and  $x_0\in \mathbb{R}$ is  a generic time coordinate. 
Let us consider a   free  EM  field  
on this background spacetime, the Lagrangian density (in vacuum) of which is given by
\begin{eqnarray}
\mathcal{L}_{\rm EM}\ =\ -\frac{1}{4}\sqrt{-g}~F_{ab}F^{ab} \ ,
\label{Lagrangian}
\end{eqnarray}
where $F_{ab}$ is the EM  field, being a covariant antisymmetric tensor of rank $2$, which can be defined in terms of the EM  potential $A_a$ by
\begin{eqnarray}
F_{ab} \ =\ \partial_{a}A_b - \partial_{b}A_a\  
\label{Maxwell}
\end{eqnarray}
and $F^{ab}=g^{ac}F_{cd}g^{db}$.
The Maxwell  equations can be written in terms of the 4-potential
$A_a$, as \cite{MTW}
\begin{eqnarray}
\square A^a - A^{b; a}_{~b}\ =\ 0\ ,
\label{Maxwell1}
\end{eqnarray}
where $\square:=g^{ab}\nabla_a\nabla_b$. Employing   the Lorenz gauge in curved spacetime, $A^{a}_{~;a}=0$, it reduces to
\begin{align}
\Box A^{a}  -  {\mathcal{R}^{a}}_{b} A^{ b } = 0\ ,
\label{Maxwell2}
\end{align}
with  $\mathcal{R}_{ab}$ being  the Ricci  tensor.
Note that  Eq.~(\ref{Maxwell2}) has  the same form of the wave equation as in flat spacetime, except that the derivatives are replaced by covariant derivatives and there is an additional term proportional to the curvature.

Henceforth, we will  study the quantum  theory of EM  fields propagating on a homogeneous background  provided by a flat FLRW model:
\begin{eqnarray}
g_{ab}dx^adx^b= -N_{x_0}^2(x_0)dx_0^2+ a^2(x_0)d\mathbf{x}^2\ .
\label{metric-class}
\end{eqnarray}
By writing the d'Alembertian in Eq.~(\ref{Maxwell2}) in terms of  the connection coefficients, we have
\begin{eqnarray}
{A^{a;b}}_{;b} &=& g^{bc}\Big[ {A^{a}}_{,b,c} + \Gamma^a_{db,c}A^d+2 \Gamma^a_{dc}{A^d}_{,b}  \nonumber \\ 
&& \quad \quad + \Gamma^a_{bd} \Gamma^d_{ce}A^e -\Gamma^d_{bc}{A^a}_{,d}- \Gamma^d_{bc} \Gamma^a_{ed}A^e \Big] . \quad \quad 
\end{eqnarray}
Computation of the connection coefficients of the metric \eqref{metric-class} in rectangular coordinates (see Appendix \ref{cf2}) shows that there are no terms involving first-order spatial derivatives of the components of the EM potential in the wave equation \eqref{Maxwell2}.
According to the remark mentioned and using the radiation gauge (in which we have $A^0=0$), the field equation for each spatial component of the vector potential $\mathbf{A}(x_0, \mathbf{x})$  lines up like
\begin{eqnarray}
\Delta^{(3)} A^{i} -\frac{a^2}{N^2} \ddot{A}^{i}+ \frac{a^2}{N^2} \left[\dfrac{\dot{N}}{N}-5 \dfrac{\dot{a}}{a} \right] \dot{A}^{i}=0\ ,
\end{eqnarray}
where  $\Delta^{(3)}$ is the Laplacian associated with the spatial Cartesian  coordinates, $i=1, 2, 3$,   
and a dot  denotes a  partial derivative with respect to $x_0$. Now we can separate $\mathbf{A}(x_0, \mathbf{x})$ as 
\begin{eqnarray}
\mathbf{A}(x_0, \mathbf{x})\ =\  \underline{\mathbf{A}}(x_0)u(\mathbf{x})
\end{eqnarray} 
with  $u(\mathbf{x})$  a solution of 
\begin{eqnarray}
\Delta^{(3)} u(\mathbf{x})= - k^2 u(\mathbf{x})\ .
\end{eqnarray} 
In other words, $u(\mathbf{x})$ is the eigenfunction of $\Delta^{(3)}$ operator with a plane wave solution, i.e., the spatial part of the EM vector potential $\mathbf{A}$ could be expanded in terms of plane waves.

Performing the Legendre transformation, one gets the electric field $E^i$ as the canonically conjugate momentum for the corresponding component of the vector potential $A_i$, on a $x_0 = const.$ slice [see  Eq. (\ref{E-i})].
Then, for the pair $(A_i,E^i)$, the classical solutions of the equation of motion (\ref{Maxwell2}) can be expanded in Fourier modes as
\begin{eqnarray}
\mathbf{A}(x_0, \mathbf{x})\   &=&\  \frac{1}{(2\pi)^{3/2}} \sum_{\mathbf{k} \in \mathcal{L}} \sum_{r}^2 \mathbf{A}_{\mathbf{k}}^{r}(x_0) e^{i \mathbf{k} \cdot \mathbf{x}}, \label{27-a} \\
\boldsymbol{\pi}(x_0, \mathbf{x})\ &  =& \  \frac{1}{(2\pi)^{3/2}} \sum_{\mathbf{k} \in \mathcal{L}} \sum_{r}^2 \boldsymbol{\pi}_{\mathbf{k}}^{r}(x_0) e^{i \mathbf{k} \cdot \mathbf{x}}. \quad \quad
\label{27}
\end{eqnarray}
The wave vector  $\mathbf{k} \in (2\pi \mathbb{Z}/\ell)^{3}$ spans a three-dimensional lattice $\mathcal{L}$, with $\mathbb{Z}$ being the set of integers \cite{DLT:2012} (see Appendix \ref{Hamiltonian}).
Notice that, since the fields are purely inhomogeneous, the zero 
$\mathbf{k}$ is excluded.

For each mode $\mathbf{k}$,
the Fourier coefficients $\mathrm{\mathbf{A}}_{\mathbf{k}}^{r}$ and 
$ \boldsymbol\pi_{\mathbf{k}}^{r}$ must evidently be vectors,
\begin{eqnarray}
\mathrm{\mathbf{A}}_\mathbf{k}^{r}\ =\  \mathrm{A}_\mathbf{k}^{r}~ \boldsymbol\epsilon^{r}_\mathbf{k} \ , \quad \quad {\rm or} \quad \quad 
\mathrm{\boldsymbol\pi}_\mathbf{k}^{r}\ =\   \mathrm{\pi}_\mathbf{k}^{r}~\boldsymbol\epsilon^{r}_\mathbf{k} \ , \quad 
\end{eqnarray}
which should satisfy the given gauge conditions.
Here, $\boldsymbol\epsilon^{r}_\mathbf{k}$ are the so-called polarization vectors \cite{Ryder}.
From the radiation gauge condition $\nabla\cdot \mathbf{A}=0$, we have that 
\begin{eqnarray}
\mathrm{\mathbf{A}}_\mathbf{k}^{r}\cdot\mathbf{k}=0\ , \quad \quad {\rm or} \quad \quad 
\boldsymbol\epsilon^{r}_\mathbf{k}\cdot\mathbf{k}=0.
\end{eqnarray}
So, for a given direction of propagation $\mathbf{k}/|\mathbf{k}|$, the polarization vectors are transverse. Moreover, they  may also be chosen to be orthonormal; $\boldsymbol\epsilon^{r}_\mathbf{k} \cdot \boldsymbol\epsilon^{r'}_\mathbf{k} = \delta_{r r'}$.
Therefore, for each Fourier coefficient, we 
have two polarization vectors; i.e., $r=1,2$.

\subsection{Hamiltonian of the EM field}

From the Lagrangian (\ref{Lagrangian}), we can write the Hamiltonian form of the EM field  propagating on the classical FLRW spacetime:
\begin{eqnarray}
\mathcal{H}_{\rm EM}\ =\ \frac{N_{x_0}}{2a^3}\Big[ \boldsymbol\pi \cdot \boldsymbol\pi + 
a^6 \sum_{i<j} F_{ij}F^{ij}\Big] .
\label{Hamiltonan-den1-new}
\end{eqnarray}
In terms of the conjugated pairs $(A_i, \pi^i)$, defined in Eq.~(\ref{E-i}),  the total Hamiltonian of the EM field is obtained as\footnote{see Eq.~(\ref{detail-A}) in Appendix \ref{Hamiltonian}.}
\begin{eqnarray}
H_{\rm EM} &=&  \int d^3x ~\mathcal{H}_{\rm EM} \notag \\
&=& \frac{N_{x_0}}{2a^3} \int d^3x ~ \Big[\boldsymbol\pi \cdot \boldsymbol\pi + a^6 \partial_i \mathbf{A} \cdot \partial^i \mathbf{A}  \Big]. \quad
\end{eqnarray}
Substituting  for $A_i$ and $\pi^i$ from (\ref{27-a}) and (\ref{27}), we find\footnote{See also Eqs.~(\ref{Hamilton-app-b1})--(\ref{Hamilton-app-b3})  for more details.}
\begin{eqnarray}
H_{\rm EM} =
 \frac{N_{x_0}}{2a^3}\sum_{\mathbf{k}} \sum_{r}^{2}
 \Big[\big(\pi_{\mathbf{k}}^{r}\big)^\ast\pi_{\mathbf{k}}^{r} + 
 k^2a^4\big(\mathrm{A}_{\mathbf{k}}^{r}\big)^\ast \mathrm{A}_{\mathbf{k}}^{r}\Big].\quad
\label{Hamilton-app1}
\end{eqnarray}

The reality condition  for the EM  field $\textbf{A}(x_0, {\mathbf{x}})$ implies  that  
not all modes $\mathrm{A}_{\mathbf{k}}^r(x_0)$ of the field are independent\footnote{cf.,  for example,  Refs.~\cite{mw, AKL:2009} for the case of a scalar field.}; 
to understand this better, let us decompose a field  mode $\mathrm{A}_{\mathbf{k}}^r(x_0)$ and its momentum $\pi_{\mathbf{k}}^{r}(x_0)$ as
\begin{eqnarray}
\mathrm{A}_{\mathbf{k}}^{r}\ &:=& \ \frac{1}{\sqrt{2}}\big(\mathrm{A}_{\mathbf{k}}^{r (1)}+i\mathrm{A}_{\mathbf{k}}^{r (2)}\big),  \label{app-phi-pi-1a} \\
\pi_{\mathbf{k}}^{r}\ &:=& \ \frac{1}{\sqrt{2}}\big(\pi_{\mathbf{k}}^{r(1)}+i\pi_{\mathbf{k}}^{r (2)}\big).  \label{app-phi-pi-1b}
\end{eqnarray}
Then, from the conditions $(\mathrm{A}_{-\mathbf{k}}^{r})^\ast=\mathrm{A}_{\mathbf{k}}^{r}$ and $(\pi_{-\mathbf{k}}^{r})^\ast=\pi_{\mathbf{k}}^{r}$, we have that
\begin{eqnarray}
&&\mathrm{A}_{-\mathbf{k}}^{r (1)}=\mathrm{A}_{\mathbf{k}}^{r (1)}\ , \quad \quad \quad \mathrm{A}_{-\mathbf{k}}^{r (2)}=-\mathrm{A}_{\mathbf{k}}^{r (2)}\ , \notag \\
&&\pi_{-\mathbf{k}}^{r(1)} = \pi_{\mathbf{k}}^{r(1)}\ , \quad \quad \quad 
\pi_{-\mathbf{k}}^{r(2)}=-\pi_{\mathbf{k}}^{r(2)}\ .
\end{eqnarray}
That is, there exist relations 
between the ``positive"   ($\mathbf{k}\in {\cal L}_+$) and ``negative"   ($-\mathbf{k}\in {\cal L}_-$) modes of the  field.
This enables one to split  the lattice ${\cal L}$, for each $\mathbf{k}=(k_1, k_2, k_3)$,  into positive and negative sectors \cite{AKL:2009}, 
\begin{eqnarray}
{\cal L}_+ &=&  \{\mathbf{k}:~ k_3>0\} \cup \{\mathbf{k}:~ k_3=0, k_2>0\} \nonumber \\
&& \quad \cup \{\mathbf{k}: ~k_3=k_2=0, k_1>0\} \ ,
\end{eqnarray}
and
\begin{eqnarray}
{\cal L}_- &=&  \{\mathbf{k}: ~k_3<0\} \cup \{\mathbf{k}:~ k_3=0, k_2<0\}  \nonumber \\
&& \quad \cup \{\mathbf{k}:~ k_3=k_2=0, k_1<0\} \notag  \\ 
&=&   \{\mathbf{k}:~ -\mathbf{k}\in {\cal L}_+\} \ ,
\end{eqnarray}
respectively.
By using this fact,  we are also able to decompose any summation over $\mathbf{k}\in {\cal L}$
into its  positive and negative parts.  
In particular, by defining new variables
$Q_{\mathbf{k}}^{r}$ and $P_{\mathbf{k}}^{r}$,
\begin{eqnarray}
Q_{\mathbf{k}}^{r}\ &:=&\ \left \{
  \begin{tabular}{cc}
  $\mathrm{A}_{\mathbf{k}}^{r (1)}$ &\quad  for \quad  $\mathbf{k} \in {\cal L}_+$\ , \\
  &\\
  $\mathrm{A}_{-\mathbf{k}}^{r (2)}$ & \quad for \quad $\mathbf{k} \in {\cal L}_-$ \ , \\
  \end{tabular} \right.   \label{def-q} \\
P_{\mathbf{k}}^{r}\ &:=&\ \left \{
  \begin{tabular}{cc}
  $\pi_{\mathbf{k}}^{r(1)}$ &\quad  for \quad  $\mathbf{k} \in {\cal L}_+$\ , \\
  &\\
  $\pi_{-\mathbf{k}}^{r(2)}$ & \quad for \quad $\mathbf{k} \in {\cal L}_-$\ ,  \\
  \end{tabular} \right.
    \label{def-p}
\end{eqnarray}
we can reexpress  the Hamiltonian (\ref{Hamilton-app1}) as
\begin{eqnarray}
H_{\rm EM}(x_0) &=&  \frac{N_{x_0}}{2a^3}\sum_{\mathbf{k}\in {\cal L}} \sum_r^2 
\Big[\big(P_{\mathbf{k}}^{r}\big)^2 + k^2a^4\big(Q_{\mathbf{k}}^{r}\big)^2\Big] \nonumber \\
&~=:&  \sum_{\mathbf{k}\in{\cal L}} \sum_{r}^2
H_{\mathbf{k}}^{(r)}(x_0)\ .
\label{Hamiltonian-SF1}
\end{eqnarray}
This equation represents the Hamiltonian  of  a set  of decoupled harmonic oscillators
defined by  conjugate pairs $(Q_{\mathbf{k}}^{r}, P_{\mathbf{k}}^{r})$.
Here, $Q_{\mathbf{k}}^{r}$ and $P_{\mathbf{k}}^{r}$ are  conjugate  variables  associated with any  $\mathbf{k}$ mode, 
satisfying the relation 
$\{Q_{\mathbf{k}}^{r}, P_{\mathbf{k}^\prime}^{r^\prime}\}=\delta_{\mathbf{k}\mathbf{k}^\prime}\delta_{rr^\prime}.$

In quantum theory, quantization of a single mode $\mathbf{k}$ (for each polarization $r$) of the vector potential, is carried out in the same way as for the quantum harmonic oscillator with the dynamical variables promoted to operators on the Hilbert space
$\mathscr{H}_{r, \mathbf{k}}=L^2( \mathbb{R}, dQ_{\mathbf{k}}^{r})$, 
as
\begin{eqnarray}
\hat{Q}_{\mathbf{k}}^{r}\psi(Q_{\mathbf{k}}^{r})=Q_{\mathbf{k}}^{r}\psi(Q_{\mathbf{k}}^{r})\ , 
\end{eqnarray}
and
\begin{eqnarray}
\hat{P}_{\mathbf{k}}^{r}\psi(Q_{\mathbf{k}}^{r})=-i\hbar\big(\partial/\partial 
Q_{\mathbf{k}}^{r}\big)\psi(Q_{\mathbf{k}}^{r}).
\end{eqnarray}
Then,  the time evolution of any state $\psi(Q_{\mathbf{k}}^{r})$ of the system is generated by the Hamiltonian operator $\hat{H}_{\mathbf{k}}^{(r)}$ via the Schr\"odinger equation:
\begin{eqnarray}
 i\hbar\partial_{x_0}\psi(x_0, Q_{\mathbf{k}}^{r}) = \frac{N_{x_0}}{2a^3} 
\Big[\big(\hat{P}_{\mathbf{k}}^{r}\big)^2 + k^2a^4\big(\hat{Q}_{\mathbf{k}}^{r}\big)^2\Big] \psi(x_0, Q_{\mathbf{k}}^{r}).   \nonumber \\
\label{Hamiltonian-SF-bquantum} 
\end{eqnarray}
In the present  study, the  wave functions of the EM modes evolve with respect to a general time coordinate $x_0$ with the  lapse   $N_{x_0}$.

Equation (\ref{Hamiltonian-SF-bquantum}) shows that, by fixing $N_{x_0}=a(\eta)$ (i.e., a conformal time coordinate $x_0=\eta$),
the evolution equation of the quantum EM field on the FLRW background reduces to the Schr\"odinger equation for the wave function  of the same EM field on a flat Minkowski space with the coordinates $(\eta, x^i)$. More precisely, the corresponding (classical) equation of motion
for each mode $Q_{\mathbf{k}}^{r}$, for a given Hamiltonian $H_{\mathbf{k}}^{(r)}(x_0)$ in Eq.~(\ref{Hamiltonian-SF1}), reads
\begin{eqnarray}
\ddot{Q}_{\mathbf{k}}^{r} + \frac{\dot{m}}{m}\dot{Q}_{\mathbf{k}}^{r}+\omega_{k}^2(x_0)Q_{\mathbf{k}}^{r}=0\ ,
\label{motion1}
\end{eqnarray}
where we have defined $m(x_0)=a^3/N_{x_0}$ and 
$\omega_{k}^2(x_0)=N_{x_0}^2k^2/a^2$.
For  $N_{x_0}=a^3(\tau)$,  Eq.~(\ref{motion1}) reduces to
\begin{eqnarray}
\ddot{Q}_{\mathbf{k}}^{r} + k^2a^4 Q_{\mathbf{k}}^{r}=0\ .
\label{motion2}
\end{eqnarray}

\section{Quantum EM field on quantized background}
\label{QG-QFT}

In this section, we briefly review   Ref.~\cite{Pawlowski:2012} by  considering a  FLRW metric (\ref{metric-class})  coupled with an irrotational dust,  as a  background matter source, and then present the LQC quantization  procedure in order to quantize the gravity-matter system.

Let us consider a background gravitational system coupled to an   {\em irrotational dust}  $T$, which is homogeneous on the spatial slices of the background metric (\ref{metric-class}), and its Lagrangian density is given by \cite{Pawlowski:2012}
\begin{equation}
\mathcal{L}_{\rm D}\ =\  -\frac{1}{2}\sqrt{-g}M\big(g^{ab}\partial_aT\partial_bT +1\big),
\end{equation}
with $M$ enforcing the gradient of the dust field to be timelike.
Then, the corresponding action for the background geometry together with a standard model matter (which denotes the EM  field  action $S_{\rm{EM}}$ for our purpose in this paper) reads
\begin{eqnarray}
S\ =\  \int d^4x  \left[\frac{\sqrt{-g}}{8\pi G}\mathcal{R} + \mathcal{L}_{\rm D}\right] + S_{\rm EM}\ .
\label{action-tot}
\end{eqnarray}
The total Hamiltonian, including the gravitational and matter sectors  is given by
\begin{eqnarray}
H\ =\ \int d^3x \big[\mathcal{H}_{\rm gr} + \mathcal{H}_{\rm D} +\mathcal{H}_{\rm EM}\big],
\label{Ham-tot}
\end{eqnarray}
where $\mathcal{H}_{\rm gr}$, $\mathcal{H}_{\rm D}$, and $\mathcal{H}_{\rm EM}$ are,  respectively, the Hamiltonian density of the gravitational, dust, and EM field sectors. The  dust contribution $\mathcal{H}_{\rm D}$ to the total Hamiltonian constraint is given by
\begin{eqnarray}
\mathcal{H}_{\rm D}\ =\ N_{x_0}\sqrt{p_T^2+q^{ab}C^D_aC^D_b} ,
\label{Ham-dust}
\end{eqnarray}
where $p_T$ is the momentum  conjugate of $T$, given by
\begin{eqnarray}
p_T \ =\  \sqrt{q}\frac{M}{N_{x_0}}\left(\dot{T}+N^a\partial_aT\right)  \ .
\label{P-C}
\end{eqnarray}
Moreover,  $C^D_a$
is the spatial diffeomorphism constraint:
\begin{eqnarray}
\quad C^D_a \ =\  -p_T\partial_a T \ .
\end{eqnarray}

For the homogeneous spatial slices, here, we have  $\partial_i T=0$, so from Eq. (\ref{P-C}),  we get $p_T=\sqrt{q}M\dot{T}/N_{x_0}$ and $C^D_a=0$. Using this condition  in the  dust Hamiltonian density (\ref{Ham-dust}),  together by  imposing the canonical time gauge fixing condition (i.e., $N_{x_0}=1$ so that   
$\dot{T}=1$ and $x_0=t=T$,  which is an obvious choice for the parametrized particle and scalar field),  we can write the total Hamiltonian constraint,  including the background gravity-dust  system together with the EM field,
on the full phase space:
\begin{eqnarray}
H  &=& \int d^3x \big[p_T + \mathcal{H}_{\rm gr}  \big]  + H_{\rm EM} \nonumber \\
 &=& \ell^3\big[p_T +  \mathcal{H}_{\rm gr} \big]  + H_{\rm EM} \ \approx 0\ .
\label{Hamiltonian-constot}
\end{eqnarray} 
At this step,  the physical Hamiltonian density can be defined as 
\begin{eqnarray}
\tilde{\mathcal{H}} := -p_T \ =\   \mathcal{H}_{\rm gr}  + \frac{H_{\rm EM}}{\ell^3} \ \cdot
\label{Phys-Hamiltonian}
\end{eqnarray}
For the FLRW metric (\ref{metric-class}), the EM Hamiltonian $H_{\rm EM}$ is given by Eq.~(\ref{Hamiltonian-SF1}), and the gravitational Hamiltonian reads  \cite{Kaminski:2008,Ashtekar:2006a}
\begin{eqnarray}
H_{\rm gr} = \int d^3x \mathcal{H}_{\rm gr} =\  \frac{3\pi G}{2\alpha}b^2|v| \ , 
\end{eqnarray}
where $b/\gamma=\dot{a}/a$ is the Hubble parameter  with  $\gamma$ being the so-called Barbero-Immirzi parameter of LQG. Moreover,  $\alpha = 2\pi \gamma \sqrt{\Delta} \ell_{\rm Pl}^2 \approx 1.35\ell_{\rm Pl}^3$ (where $\Delta$ is the so-called ``area gap" given by $\Delta=3\sqrt{3}\pi \gamma \ell_{\rm Pl}^2$),
and $v= \ell^3a^3/\alpha$ is the {\em oriented volume}  with $b$ being its conjugate momentum satisfying $\{v, b\}=2$.

Following the Dirac quantization scheme for the constrained systems, 
a total kinematical Hilbert space  for the above gravity-matter system  can be defined as $\mathscr{H}_{\rm kin} = \mathscr{H}_{\rm gr} \otimes \mathscr{H}_{T}  \otimes \mathscr{H}_{\rm EM}$, where
the matter sectors are
quantized according to the Schr\"odinger picture with the Hilbert spaces
 $\mathscr{H}_{T}=L^2 (\mathbb{R}, dT)$ and $\mathscr{H}_{r, \mathbf{k}}=L^2( \mathbb{R}, dQ_{\mathbf{k}}^{r})$ (for each mode), and gravity is quantized due to LQC with $\mathscr{H}_{\rm gr}=L^2 (\bar{\mathbb{R}}, d\mu_{\rm Bohr})$ (in which $\bar{\mathbb{R}}$ is the Bohr compactification of the real line and $d\mu_{\rm Bohr}$ is the Haar measure on it \cite{Ashtekar:2003}). Now, from Eq.~(\ref{Phys-Hamiltonian}), the corresponding quantum operators  on $\mathscr{H}_{\rm kin}$ are those  acting on physical states $\Psi(v, Q, T)\in \mathscr{H}_{\rm kin}$ such that 
\begin{eqnarray}
i\hbar \partial_T \Psi(v, Q, T) =  \left[\hat{\mathcal{H}}_{\rm gr}  + \hat{H}_{\rm EM}/\ell^3\right]  \Psi(v, Q, T). \quad
\label{QG-Shroedinger}
\end{eqnarray}
Here, $v$ is the eigenstate of the volume operator of the quantum background geometry,
\begin{eqnarray}
\hat{v}  |v\rangle = v |v\rangle \quad {\rm and} \quad \hat{\mathcal{N}} |v\rangle = |v + 1\rangle\ ,
\end{eqnarray}
where $\hat{\mathcal{N}}\equiv \widehat{\exp(ib/2)}$ is the quantum operator corresponding to the exponential version of $b$.
Moreover, $\{ |v\rangle \}$ is the basis of eigenstates of the volume operator $\hat{v}$ satisfying  
$\langle v |v^\prime \rangle=\delta_{v, v^\prime}$.
On this quantum background,  
the gravitational Hamiltonian constraint becomes \cite{Kaminski:2008, Pawlowski:2012}
\begin{eqnarray}
\hat{H}_{\rm gr}  \ =\  - \frac{3\pi G}{8\alpha}\sqrt{|\hat{v}|}\big(\hat{\mathcal{N}}-\hat{\mathcal{N}}^{-1}\big)^2\sqrt{|\hat{v}|}\ .
\end{eqnarray} 
In the absence of external fields, physical states must satisfy 
\begin{eqnarray}
i\hbar \partial_T \Psi_o(v, T) \ =\  \hat{\mathcal{H}}_{\rm gr} \Psi_o(v, T)\ ,
\label{QG-Shroedinger-geo}
\end{eqnarray}
where we have replaced  $\Psi$ by  the pure gravitational states $\Psi_o\in\mathscr{H}_{\rm gr}$ only. These  states provide the scalar product  with a finite norm
\begin{eqnarray}
\langle \Psi_o | \Psi_o^\prime \rangle \ =\  \sum_v \Psi_o^\ast(v, T_0)\Psi_o^\prime(v, T_0)\ ,
\label{norm-gravity}
\end{eqnarray}
and belong to the physical Hilbert space $\mathscr{H}_{\rm Phys}^o$ of the geometry. Here, $T_0$ is any ``instant" of relational time $T$.  We will set  the norm 
of the scalar product of the gravitational states to be $\langle \Psi_o | \Psi_o \rangle=1$.

As we can see from Eq.~(\ref{QG-Shroedinger}), 
the background geometry is quantized due to quantum gravity, so 
the classical {\em geometrical} variables  in the matter sector 
(\ref{Hamiltonian-SF1})
should be replaced by the corresponding quantum operators on 
$\mathscr{H}_{\rm gr}$. 
Therefore, $\hat{H}_{\rm EM}$ in Eq.~(\ref{QG-Shroedinger}) can be written as
\begin{eqnarray}
\hat{H}_{\rm EM}/\ell^3 
\ =:\  \sum_{\mathbf{k}\in{\cal L}} \sum_{r}^2
\hat{H}_{T, \mathbf{k}}^{(r)}\ , \quad\quad
\label{Hamiltonian-SF1-QG}
\end{eqnarray} 
where
\begin{eqnarray} 
\hat{H}_{T, \mathbf{k}}^{(r)} :=  \frac{1}{2\ell^3} 
\Big[\widehat{a^{-3}}\otimes \big(\hat{P}_{\mathbf{k}}^{r}\big)^2 + k^2\hat{a} \otimes \big(\hat{Q}_{\mathbf{k}}^{r}\big)^2\Big].~ \quad
\label{Hamiltonian-SF1-QG2}
\end{eqnarray}
Then,  from Eqs.~(\ref{Hamiltonian-SF1-QG}) and (\ref{QG-Shroedinger}),  
the Schr\"odinger equation 
for each mode and polarization of the EM field  is  written as
\begin{eqnarray}
i\hbar \partial_T \Psi(v, Q, T)   & =&     \Big[\hat{\mathcal{H}}_{\rm gr}  + \hat{H}_{T, \mathbf{k}}^{(r)}  \Big] \Psi(v, Q, T)  \nonumber \\
&=:& \hat{\tilde{\mathcal{H}}}_{\mathbf{k}} \Psi(v, Q, T) \ ,
\label{QG-Shroedinger-2}
\end{eqnarray}
where $\hat{\tilde{\mathcal{H}}}_{\mathbf{k}}$ is the physical Hamiltonian for each mode.

\section{Emerging geometries}

For the test field propagating on the quantized background provided in the previous section, we  apply  two approximations in order to predict the emerging effective background geometry: (i) the test field approximation in which backreaction is discarded (see Sec.~\ref{Test-field} below), and (ii) the Born-Oppenheimer  approximation  in which  the  backreaction  is  present (see  Sec.~\ref{BO-approx} below). Then,  we  discuss  the phenomenological features of those two geometries (see Sec.~\ref{Phenomenology}).

\subsection{Test-field approximation}
\label{Test-field}

The evolution equation (\ref{QG-Shroedinger-2}) is rather analogous to the Schr\"odinger equation (\ref{Hamiltonian-SF-bquantum}). However,  the quantum geometry operators in $\hat{\mathcal{H}}_{\rm gr}$ do not depend on time, and Eq.~(\ref{QG-Shroedinger-2}) provides quantum evolution for the state $\Psi(v, Q, T)$ that depends on the $\mathbf{k}$th mode of the test EM field and the quantum geometry, while  Eq.~(\ref{Hamiltonian-SF-bquantum}) evolves the state $\psi(Q, T)$ (of the EM test field only) on the (time-dependent) classical FLRW  background. To make two evolutions comparable, we should work in the ``interaction picture" \cite{AKL:2009} by setting
\begin{eqnarray}
\Psi(v, Q, T) \ =\  e^{-(i/\hbar)\hat{H}_{o}(T-T_0)}\Psi_{\rm int}(v, Q, T) 
\end{eqnarray}
for any instant of relational time $T_0$. Then, Eq.~(\ref{QG-Shroedinger-2}) reduces to
\begin{eqnarray}
i\hbar \partial_T \Psi_{\rm int}   &=&
\frac{1}{2\ell^3}  \Big[\widehat{a^{-3}}(T)\otimes \big(\hat{P}_{\mathbf{k}}^{r}\big)^2 \nonumber \\
&& \quad \quad\quad    + k^2 \hat{a}(T) \otimes \big(\hat{Q}_{\mathbf{k}}^{r}\big)^2\Big]   \Psi_{\rm int}\ ,  \quad\quad 
\label{QG-Shroedinger-2int}
\end{eqnarray}
where any operator $\hat{A}(T)$ is the corresponding time-dependent  operator in the Heisenberg picture now:
\begin{eqnarray}
\hat{A}(T) = e^{\frac{i}{\hbar}\hat{H}_o(T-T_0)}\hat{A}e^{-\frac{i}{\hbar}\hat{H}_o(T-T_0)}\ .
\end{eqnarray}
In this picture, the quantum geometry (presented by the heavy degrees of freedom $\hat{\mathcal{H}}_{\rm gr} $ in the total Hamiltonian) is in effect described in the Heisenberg picture (i.e., its states are frozen at  time $T_0$,  but the scale factor operators evolve in time), while the test field (as a perturbation denoted as light degrees of freedom, provided by $\hat{H}_{T, \mathbf{k}}^{(r)}$) is described using the Schr\"odinger picture.

When a test-field approximation is considered, i.e., when a matter backreaction is discarded, then, for any time $T$,  one can decompose the total   wave function 
as \cite{AKL:2009}
\begin{eqnarray}
\Psi_{\rm int}(v, Q, T)\ =\  \Psi_o(v, T_0)\otimes\psi(Q, T)\ 
\end{eqnarray}
so that  the geometry's quantum state $\Psi_o(v, T)$ obeys the ``unperturbed"  evolution equation
$i\hbar\partial_{T}\Psi_o(v, T)=\hat{\mathcal{H}}_{\rm gr}\Psi_o(v, T)$.
In this case, the Schr\"odinger equation (\ref{QG-Shroedinger-2})
reduces to an evolution equation for $\psi(Q, T)$  only:
\begin{eqnarray}
i\hbar \partial_T  \psi  =
\frac{1}{2\ell^3}  \Big[\big\langle\widehat{a^{-3}}(T)\big\rangle_o\big(\hat{P}_{\mathbf{k}}^{r}\big)^2    + k^2\big\langle\hat{a}(T)\big\rangle_o  \big(\hat{Q}_{\mathbf{k}}^{r}\big)^2\Big]   \psi .  \quad  
\label{evol-eq1}
\end{eqnarray}
By comparison with Eq.~(\ref{Hamiltonian-SF-bquantum}), Eq.~(\ref{evol-eq1}) can be seen as
a(n) (Schr\"odinger) evolution equation for the $\mathbf{k}$th mode of the EM field on a dressed spacetime with 
$(M, \bar{g}_{ab})$,
\begin{eqnarray}
\bar{g}_{ab}dx^adx^b\ =\ - \bar{N}_T^2(T)dT^2 + \bar{a}^2(T)d\mathbf{x}^2\ ,
\label{dressed-metric}
\end{eqnarray}
where the metric components $\bar{N}_T$ and $\bar{a}$  have relations with the expectation values of the original spacetime operators as
\begin{eqnarray}
\bar{N}_T/\bar{a}^3 = \ell^{-3} \big\langle\widehat{a^{-3}}(T)\big\rangle_o\ , \quad   \quad
\bar{N}_T\bar{a} = \ell^{-3} \big\langle\hat{a}(T)\big\rangle_o .\quad
\label{dressed-metric-eq}
\end{eqnarray}
Equation (\ref{dressed-metric-eq}) provides two  equations for two unknown variables of the emergent dressed background  metric (\ref{dressed-metric}).
Using these two equations, we find that
\begin{eqnarray}
\bar{N}_T(T) &=&  \ell^{-3} \left[\big\langle\widehat{a^{-3}}(T)\big\rangle_o ~\big\langle\hat{a}(T)\big\rangle_o^3 \right]^{\frac{1}{4}}, \nonumber \\
  \bar{a}(T) &=& \left[\frac{\big\langle\hat{a}(T)\big\rangle_o}{\big\langle\widehat{a^{-3}}(T)\big\rangle_o}\right]^{\frac{1}{4}}. \quad
\label{dressed-metric-eq2}
\end{eqnarray}

The emerging  geometry (\ref{dressed-metric}), provided   by the components (\ref{dressed-metric-eq2}), does not depend on any specific chosen mode of the test EM field, and therefore, there is no  violation of the Lorentz symmetry.  However, for a massive field (such as the Proca field), one may expect a  mode-dependent  solution  such that a rainbow metric can emerge for the effective background geometry \cite{Andrea:2015}.

\subsection{Born-Oppenheimer approximation}
\label{BO-approx}

So far, we have seen that, when the backreaction of the EM field on the background quantum  geometry is discarded due to a test-field approximation procedure, no deviation from the local Lorentz symmetry emerges,  and the dressed background metric is independent of the chosen EM field mode. However, it is possible to  lift this condition by employing  a Born-Oppenheimer approximation  scheme in the evolution equation (\ref{QG-Shroedinger-2int}). 

The Born-Oppenheimer approximation is a particular form of a general method called
the  {\em adiabatic} approximation,  which underlines the theory of molecular motion in atomic physics, based on a general  assumption  that  the heavy degrees of freedom (such as nuclei) evolve so slowly compared with the light degrees of freedom (i.e., the electron in the atom) that the force between the atoms can still be calculated by differentiation of the energies $E_{e}^{(l)}$ of each perturbation's (i.e., electronic) eigenstates $l$.  The function $E_{e}^{(l)}$
thus serves as the potential energy for the evolution of the nuclei if the electrons  are in state $l$. An extension of this approximation has also been made in the context of quantum gravity, especially in the Wheeler-deWitt (see Ref.~\cite{Kiefer:2004} and references therein) and LQG approaches \cite{Rovelli:2008, Giesel:2009}.

As we have seen before, 
the dynamics of the gravity-field system is generated by the physical Hamiltonian (\ref{QG-Shroedinger-2}).
Here,  the gravitational  sector of the  Hamiltonian (\ref{QG-Shroedinger-2})
represents  the heavy degrees of freedom, and the field Hamiltonian represents  the light degrees of freedom. One difference to the Hamiltonian in a hydrogen atom is that, here,  the light degrees of freedom (i.e., the EM field) are not only the functions of field variable $(Q, P)$  but are functions of 
the gravitational variable $\hat{v}$ as well. Our task will be then to obtain solutions to the full eigenvalue problem,
\begin{eqnarray}
 \Big\{ \hat{\mathcal{H}}_{\rm gr} 
&+& \frac{1}{2\ell^3}\Big[\widehat{a^{-3}}\otimes \big(\hat{P}_{\mathbf{k}}^{r}\big)^2 \nonumber \\
&+&  k^2\hat{a} \otimes \big(\hat{Q}_{\mathbf{k}}^{r}\big)^2\Big] \Big\}\tilde{\Psi}(v, Q)  = E_\mathbf{k} \tilde{\Psi}(v, Q), \quad \quad
\label{QG-Shroedinger-2BO}
\end{eqnarray}
where $E_\mathbf{k}$ is the total energy eigenvalue of each mode.

The Born-Oppenheimer approximation  consists of assuming that the solution to (\ref{QG-Shroedinger-2BO})  has the form
\begin{eqnarray}
\tilde{\Psi}(v, Q) &=&  \sum_{\mu, l} c_{\mu l} \xi^\mu_l(v) \otimes \chi^{l} (v, Q) , 
\label{BO1}
\end{eqnarray}
for the stationary states $\tilde{\Psi}\in \mathscr{H}~(=\mathscr{H}_{\rm gr}\otimes \mathscr{H}_{r, \mathbf{k}})$, where $c_{\mu l}$ are  expansion  coefficients. 
For the wave function (\ref{BO1}), the Schr\"odinger equation (\ref{QG-Shroedinger-2}) is separated into a stationary state equation for the field mode,
\begin{eqnarray}
\hat{H}_{T, \mathbf{k}}^{(r)}~ \chi^{l}(Q; v)  = \epsilon^{(l)}_{\mathbf{k}}(\hat{v}) ~ \chi^{l}(Q; v) \ ,
\label{Eigenvalue-field1}
\end{eqnarray}
on the specified (fixed) background quantum geometry\footnote{That is, the dependence on gravitational variable $v$ has to be understood in a parametrical way so that we assume that $v$
is fixed and regard $\chi^{l}(Q; v)$ as an element on $\mathscr{H}_{r, \mathbf{k}}$. In a sense, this means we can solve Eq.~(\ref{Eigenvalue-field1}) for each external parameter $Q$ separately.} 
with the state $\xi^\mu_l(v)$, 
and a  second equation describing the evolution of the background:
\begin{eqnarray}
\big[\hat{\mathcal{H}}_{\rm gr} +  \lambda \epsilon^{(l)}_{\mathbf{k}}(\hat{v})\big] \xi^\mu_l(v) =  E^\mu_l \xi^\mu_l(v)\ .
\label{Eigenvalue-grav1}
\end{eqnarray}
Here, $\lambda$ is a real  constant  introduced to
keep track of the number of times the perturbation enters, 
and  $\epsilon^{(l)}_{\mathbf{k}}(\hat{v})$ is the eigenvalue of the $l$th eigenstate of the field mode
(light variable),  but  it is still an operator on the quantum geometrical Hilbert space. Now, due to the orthonormality of  $\chi^l$, we get
\begin{eqnarray}
\epsilon^{(l)}_{\mathbf{k}}(\hat{v}) &=& \frac{1}{2\ell^3} 
\Big[\widehat{a^{-3}} \big\langle \chi^{l} \big| \big(\hat{P}_{\mathbf{k}}^{r}\big)^2 \big| \chi^{l} \big\rangle  \nonumber \\
&& \quad \quad \quad + k^2\hat{a} \big\langle \chi^{l}\big| \big(\hat{Q}_{\mathbf{k}}^{r}\big)^2 \big| \chi^{l} \big\rangle \Big]. \quad \quad
\end{eqnarray}
Note that $\xi^\mu_l$ is the  eigenstate of the (backreacted) geometry due to the presence of the $l$th eigenstate of the EM field, which is different from  unperturbed eigenstates $\xi^\mu_o$.
If $\lambda\epsilon^{(l)}_{\mathbf{k}}$ in Eq.~(\ref{Eigenvalue-grav1}) were negligible,  then the  Born-Oppenheimer description here reduces to the test field approximation, in which the gravitational sector does not know anything about the field modes and the gravitational wave function is simply given by the solution to $\hat{\mathcal{H}}_{\rm gr} \xi_o^\mu(v) = E^\mu_o \xi_o^\mu(v)$ for which  $E^\mu_l=E^\mu_o$.
However, in the case $\lambda\epsilon^{(l)}_{\mathbf{k}}\neq 0$ here, the field mode influences the heavy degrees of freedom {\em effectively}, and $\xi^\mu_l\neq \xi^\mu_o$ (although it is not taken into account with respect to full dynamics).

After analyzing the eigenvalue equation (\ref{Eigenvalue-field1}) for the field mode, we can find the eigenfunctions $\chi^{l}(Q; v)$ and eigenvalues  $\epsilon^{(l)}_{\mathbf{k}}(\hat{v})$. Then, by substituting these functions in Eq.~(\ref{Eigenvalue-grav1}), we can compute the coefficients $\xi^\mu_l(v)$, and thus  we will find the complete solutions (\ref{BO1}) to the full quantum theory of EM modes on the quantum FLRW background. 
To do so, we employ  {\em perturbation theory} in order to compute 
approximate solution for the eigenstates $\xi^\mu_l$. In this mechanism, since $\{\xi^\mu_o\}$ form a complete set of bases for the unperturbed wave function $\Psi_o$ (with the eigenvalues $E^\mu_o$),  we can expand the perturbed states $\xi^\mu_l$ in terms of them as well, 
\begin{eqnarray}
\xi^\mu_l \ =\  N(n)\Big[\xi^\mu_o +  \sum_{\nu\neq \mu} \sum_n  \lambda^n \beta_{\mu \nu}^{n (l)}\xi^\nu_o\Big],
\label{eigenstate-pert1}
\end{eqnarray}
in which $\beta_{\mu\nu}^{n (l)}$ (with $n\neq 0$) are the expansion coefficients for the $n$th  order  of perturbation. By substituting (\ref{eigenstate-pert1}) in the Schr\"odinger equation (\ref{Eigenvalue-grav1}),
we can find the coefficients $\beta_{\mu\nu}^{n (l)}$ and the energy eigenvalues $E^\mu_l=E^\mu_o + \lambda E^{\mu(1)}_{l} + \lambda^2 E^{\mu(2)}_{l} + \cdot\cdot\cdot$.
To the first order of perturbation (i.e., for $n=1$), we get
\begin{eqnarray}
\beta_{\mu\nu}^{1 (l)}\ :=\ \frac{\epsilon_{\mu\nu}^{(l)}(k)}{E_o^\mu - E_o^\nu}\ , 
\end{eqnarray}
where
\begin{eqnarray}
\epsilon_{\mu\nu}^{(l)} &:=& \big\langle \xi_o^\nu \big| \epsilon^{(l)}_{\mathbf{k}}(\hat{v})  \big|\xi^\mu_o  \big\rangle \nonumber \\
 &=& \frac{1}{2\ell^3} 
\Big[\big\langle \xi_o^\nu \big|\widehat{a^{-3}}  \big|\xi^\mu_o  \big\rangle \big\langle \chi^{l} \big| \big(\hat{P}_{\mathbf{k}}^{r}\big)^2| \chi^{l} \big\rangle  \nonumber \\
&& \quad \quad \quad + k^2\big\langle \xi_o^\nu \big| \hat{a}  \big|\xi^\mu_o  \big\rangle \big\langle \chi^{l} \big| \big(\hat{Q}_{\mathbf{k}}^{r}\big)^2 \big| \chi^{l} \big\rangle \Big] \nonumber \\
&=:& A_{\mu\nu}^{(l)} + k^2 B_{\mu\nu}^{(l)}\ .
\label{epsilon-BO}
\end{eqnarray}
Moreover,   the normalization coefficient $N(n)$ is chosen such that $N(0)=1$ and $\lambda = 0$ for the unperturbed case. The first-order corrections to the energy are then given by
\begin{eqnarray}
E^{\mu}_l\ =\  E^\mu_o + \lambda \big\langle \xi_o^\mu \big| \epsilon^{(l)}_{\mathbf{k}}(\hat{v})  \big|\xi^\mu_o  \big\rangle + \mathcal{O}(\lambda^2)\ .
\end{eqnarray}
By setting $\langle \xi^\mu_l | \xi^\mu_l \rangle=1$, we obtain $N(1)=1$ up to the first-order corrections. Then, 
$\xi^\mu_l$ becomes
\begin{eqnarray}
\xi^\mu_l \ =\  \xi^\mu_o + \lambda \sum_{\nu\neq \mu}  \beta_{\mu\nu}^{1 (l)} \xi^\nu_o\ .
\label{BO-Pert2}
\end{eqnarray}

To have  better insight regarding the evolution of  the  field modes on the herein quantum background, when the coefficients $\xi^\mu_l$ are already known  through  Eq.~(\ref{BO-Pert2}), similar to  the  case we had  in  the test-field approximation, we will describe the situation 
in a way in which  the background geometry can be treated almost classically. 
By substituting Eq.~(\ref{BO-Pert2}) in the total wave function (\ref{BO1}), we obtain
\begin{eqnarray}
\tilde{\Psi}(v, Q) &=&  \sum_{\mu, l}\Big[ c_{\mu}^o \xi^\mu_o \otimes b_l \chi^{l}  + \lambda 
\sum_{\nu\neq \mu} c_{\mu l}  \beta_{\mu\nu}^{1 (l)} \xi^\nu_o \otimes \chi^{l}  \Big] \nonumber \\
&=:& \Psi_o(v)\otimes \psi(Q) + \delta \Psi(v, Q) ,
\label{BO1-Perturbation}
\end{eqnarray}
in which  the first term (on the right-hand side) denotes the unperturbed wave function  and the second term denotes the influence of the light degrees of freedom (i.e., EM mode perturbation) on the geometry quantum state.
In particular,  we are interested in a description of quantum modes on a(n) (effective) classical background,  so we will further assume that the field eigenstates $\chi^l$  are not affected by quantum geometry, i.e.,  they are not entangled with the background eigenstates $\xi^\mu_l$. Then, we can decompose  the perturbation wave function as\footnote{
The backreaction term in Eq.~(\ref{BO1-Perturbation}) can be decomposed as
\begin{eqnarray}
\delta\Psi(v, Q) &=& \lambda \sum_{\mu, l} c_{\mu l}  \sum_{\nu\neq \mu}   \beta_{\mu\nu}^{1 (l)} \xi^\nu_o \otimes \chi^{l} \nonumber \\
&=:& \lambda \sum_{\mu}  \sum_{\nu\neq \mu} \tilde{c}_{\mu}^o  \beta_{\mu\nu}^{1}   \xi^\nu_o \otimes \sum_{l}  \tilde{b}_{l} \chi^{l}\ .
\nonumber
\end{eqnarray}
Notice that the last relation above was obtained by assuming that $c_{\mu l} = \tilde{c}_{\mu}^o \tilde{b}_{l}$, and $\beta_{\mu \nu}^{1 (l)}$ is  the same for all $\chi^l$, such that $\beta_{\mu \nu}^{1 (l)}\equiv\beta_{\mu \nu}^{1}$. In other words, decomposition (\ref{BO1-Perturbation-2}) is only relevant when the field eigenstates are degenerate.
}
\begin{eqnarray}
\delta\Psi(v, Q)\ \approx\  \delta\Psi_o(v)\otimes\psi(Q) \ , \quad
\label{BO1-Perturbation-2}
\end{eqnarray}
where
\begin{eqnarray}
\delta\Psi_o(v)\ =\ \lambda \sum_{\mu} \sum_{\nu \neq \mu} c_{\mu}^o  \beta_{\mu \nu}^{1} \xi^\nu_o~.  \quad \quad
\end{eqnarray}
Now, by substituting the total (perturbed) wave function (\ref{BO1-Perturbation}), with  the disentangled perturbation term $\delta\Psi=\delta\Psi_o\otimes\psi$ above, into the total time-dependent Schr\"odinger equation (\ref{QG-Shroedinger-2}), we obtain
 \begin{eqnarray}
i\hbar \partial_T \psi(Q, T)   & =&   \frac{1}{2\ell^3}\Big[\Big(\langle \widehat{a^{-3}} \rangle_o + \lambda \langle \widehat{a^{-3}} \rangle_\delta\Big) \big(\hat{P}_{\mathbf{k}}^{r}\big)^2   \nonumber \\
&& \quad \quad + k^2 \Big(\langle \hat{a} \rangle_o + \lambda \langle \hat{a} \rangle_\delta\Big)  \big(\hat{Q}_{\mathbf{k}}^{r}\big)^2\Big] \psi . \quad \quad \quad \
\label{QG-Shroedinger-2-BO}
\end{eqnarray}
Here, we have defined $\langle \widehat{a^{-3}} \rangle_\delta\equiv \langle \Psi_o | \widehat{a^{-3}} |\delta \Psi_o\rangle$ and $\langle \hat{a} \rangle_\delta\equiv \langle \Psi_o | \hat{a} |\delta \Psi_o\rangle$. We have further used the fact  that  the perturbed wave function $\delta\Psi_o$ is orthogonal to $\Psi_o$, so $\langle \Psi_o|\delta\Psi_o\rangle=0$.
To make a comparison between the  backreacted  quantum geometry implemented by Eq.~(\ref{QG-Shroedinger-2-BO}) and the dressed  background (\ref{evol-eq1}) provided by the test-field approximation, it is convenient to  expand correction terms in Eq.~(\ref{QG-Shroedinger-2-BO}),   in terms of the unperturbed background eigenstates $\xi^\mu_o$, as
\begin{eqnarray}
\langle \widehat{a^{-3}} \rangle_\delta &=&  \sum_{\sigma, \mu} \sum_{\nu \neq \mu}(c_{\sigma}^{o})^\ast c_{\mu}^o \frac{\langle \xi^\sigma_o | \widehat{a^{-3}} |\xi^\nu_o \rangle}{E_o^\mu - E_o^\nu} \epsilon_{\mu\nu}^{(l)}(k) , \quad  \ 
\label{delta-1}
 \\
 \langle \hat{a} \rangle_\delta &=& \sum_{\sigma, \mu} \sum_{\nu \neq \mu}(c_{\sigma}^{o})^\ast c_{\mu}^o  \frac{\langle \xi^\sigma_o | \hat{a} |\xi^\nu_o \rangle}{E_o^\mu - E_o^\nu} \epsilon_{\mu\nu}^{(l)}(k) \  .
\label{delta-2}
\quad
\end{eqnarray}
Since $\langle \widehat{a^{-3}} \rangle_\delta$ and $\langle \hat{a} \rangle_\delta$ are functions of  
$\epsilon_{\mu\nu}(k)$,   these coefficients  depend on  the mode $\mathbf{k}$ of the EM field.
Now, as before, by comparison with Eq.~(\ref{Hamiltonian-SF-bquantum}), 
Eq.~(\ref{QG-Shroedinger-2-BO}) can be seen as
an  evolution equation for  $\mathbf{k}$th mode  of the EM field on a dressed spacetime with 
$(M, \tilde{g}_{ab})$,
\begin{eqnarray}
\tilde{g}_{ab}dx^adx^b\ =\ - \tilde{N}_T^2(T)dT^2 + \tilde{a}^2(T)d\mathbf{x}^2\ ,
\label{dressed-metric-BO}
\end{eqnarray}
where  $\tilde{N}_T$ and $\tilde{a}$  are related to   the expectation values of the original spacetime operators as
\begin{eqnarray}
\tilde{N}_T/\tilde{a}^3 &=& \ell^{-3} \big[ \big\langle\widehat{a^{-3}}\big\rangle_o +\lambda \langle \widehat{a^{-3}} \rangle_\delta\ \big],  \\
\tilde{N}_T\tilde{a} &=& \ell^{-3} \big[ \big\langle\hat{a}\big\rangle_o + \lambda \langle \hat{a} \rangle_\delta\ \big] .
\end{eqnarray}
Solving the  equations above, we obtain
\begin{eqnarray}
\tilde{N}_T &=&  \ell^{-3} \Big[\Big(\big\langle\widehat{a^{-3}}\big\rangle_o + \lambda\langle \widehat{a^{-3}} \rangle_\delta\Big) \notag \\
&& \quad \quad \quad \quad \times \Big(\big\langle\hat{a}\big\rangle_o + \lambda \langle \hat{a} \rangle_\delta\Big)^3 \Big]^{\frac{1}{4}}, \quad \quad \quad  \\
  \tilde{a}(T) &=& \left[\frac{\big\langle\hat{a}\big\rangle_o + \lambda \langle \hat{a} \rangle_\delta}{\big\langle\widehat{a^{-3}}\big\rangle_o + \lambda \langle \widehat{a^{-3}} \rangle_\delta}\right]^{\frac{1}{4}}. \quad
\label{dressed-metric-eq2-BO}
\end{eqnarray}
By expanding the relations above, we can rewrite them as 
\begin{eqnarray}
\tilde{N}_T &=&    \bar{N}_T\big[1 + \lambda \beta(k)\big]  ,\\
\tilde{a}(T) &=&   \bar{a}(T)\big[1 + \lambda \sigma(k)\big] ,  \quad\quad
\label{dressed-metric-eq2-BO2}
\end{eqnarray}
where $\bar{N}_T$ and $\bar{a}$ are given by Eq.~(\ref{dressed-metric-eq2}) and $\beta(k)$ and $\sigma(k)$ are defined as
\begin{eqnarray}
\beta(k)  &=& \frac{1}{4}\frac{\langle \widehat{a^{-3}} \rangle_\delta}{\big\langle\widehat{a^{-3}}\big\rangle_o} + \frac{3}{4}\frac{\langle \hat{a} \rangle_\delta}{\big\langle\hat{a}\big\rangle_o} + \cdot\cdot\cdot \ , \\
\sigma(k) &=& \frac{1}{4}\frac{\langle \hat{a} \rangle_\delta}{\big\langle\hat{a}\big\rangle_o} -\frac{1}{4}\frac{\langle \widehat{a^{-3}} \rangle_\delta}{\big\langle\widehat{a^{-3}}\big\rangle_o} + \cdot\cdot\cdot  \ .
\end{eqnarray}
This indicates that in the presence of the backreaction
the components  of the effective metric (\ref{dressed-metric-BO}) depend on the mode of the field. Thus, each mode of the EM field probes a  dressed background spacetime differently from the other modes, so the effective background looks like a rainbow metric (see, for example, Refs.~\cite{Lafrance:1995,Magueijo:2004}) from the   field's point of view.

\subsection{Rainbow metric: Violation of the local Lorentz symmetry}
\label{Phenomenology}

So far, we have seen that when the energy $\epsilon_{\mathbf{k}}^{(l)}$ of the mode's eigenstate $\chi^l$ is high  the effective background probed by the mode becomes $k$ dependent, due to the backreaction effects predicted  by the Born-Openheimer approximation. 
Now, we provide an interpretation for  a ``classical observer"  measuring  the background metric  $\bar{g}_{ab}$  probed by low-energy modes [see Eq.~(\ref{dressed-metric})], while the high-energy modes propagate on $\tilde{g}_{ab}(k)$ [given by Eq.~(\ref{dressed-metric-BO})].

In general, a (classical) cosmological  observer with a normalized 4-velocity $u^a=(1/\bar{N}_T, 0, 0 ,0)$  measures the energy of a  particle with the 4-momentum $k_a=( k_0, k_1, k_2, k_3) $ to be
$E = k_a u^a$. 
The normalization condition for the 4-velocity implies $\bar{g}_{ab}u^au^b=-1$. 
For the emerging background (\ref{dressed-metric-BO}), the on-shell relation for the photon with the mass $m_0=0$ becomes $\tilde{g}^{ab}k_a k_b  = 0$  so that
\begin{eqnarray}
\tilde{\omega}^2_k(k) &=&  k_0^2\ =\  \dfrac{\bar{N}_T^2}{\bar{a}^2} k^2 \left(\frac{1+ \lambda \beta(k)}{1+ \lambda \sigma(k)}\right)^2 \nonumber \\
&=:& f^2(k)\bar{\omega}^2_k(k). \quad \quad
\label{dispersion-BO}
\end{eqnarray}
This indicates that from the point of view of the classical cosmological observer on the backreacted background (\ref{dressed-metric-BO}) the  energy  $\bar{\omega}_k$ of the EM modes  is modified by a  mode-dependent function,  $f(k)\neq1$.

The appropriate rescaled components of the physical momentum $p$ and the energy $E$ in the tetrad frame of the classical  observer (where $\bar{g}_{ab}=\eta_{AB}e_{a}^{A} e^{B}_{b}$, with the internal indices $A, B=0, 1, 2, 3$ and the internal metric $\eta_{AB}$ such that $\eta^{AB} = e^{A a} e^{B}_{a}$) are given by
\begin{eqnarray}
k_{\hat{0}}\ = \  \frac{\tilde{\omega}}{\bar{N}_T}\ =:\ E  , \quad\quad {\rm and} \quad\quad k_{I}\ =\ \frac{k_{i}}{\bar{a}} ~, 
\end{eqnarray}
where $k_{A}=e_{A}^{a} k_a $, $p^2= k_{I} k^{I}$  (with $\hat{0}$ denoting the zeroth component of the internal metric and  $I, J=1, 2, 3$ being the three-dimensional internal  indices). 
Then, the 3-velocity $v^{I}$ of the photon measured in the three-dimensional internal basis  of the cosmological  observer reads
\begin{eqnarray}
v^{I} = \frac{dE}{dk_{I}}\ ,
\end{eqnarray}
with the squared norm $|v|^2=v_Iv^I$, such that 
\begin{eqnarray}
|v|  &\approx& 1+  \lambda \big(L_1^{(l)}+3L_2^{(l)}p^2\big)+\mathcal{O}(p^4,\lambda^2). \quad  \quad
\label{speed-photon}
\end{eqnarray}
Here, $L_1^{(l)}$ and $L_2^{(l)}$ are  functions of quantum fluctuation of the background geometry and the field mode, 
\begin{eqnarray}
L_1^{(l)} &=& \frac{1}{2} \sum_{\sigma, \mu} \sum_{\nu\neq \mu} \frac{(c_\sigma^o)^\ast c_\mu^o Z_{\sigma\nu}}{E_o^\mu - E_o^\nu} A_{\mu\nu}^{(l)}\ , \quad   \\
L_2^{(l)} &=& \frac{\bar{a}^2}{2}  \sum_{\sigma, \mu} \sum_{\nu\neq \mu} \frac{(c_\sigma^o)^\ast c_\mu^o Z_{\sigma\nu}}{E_o^\mu - E_o^\nu} B_{\mu\nu}^{(l)} \ , \ \quad  
\quad
\end{eqnarray}
where
\begin{eqnarray}
Z_{\sigma\nu}\ :=\ \left[\frac{\langle \xi^\sigma_o | \widehat{a^{-3}} |\xi^\nu_o \rangle}{\langle \widehat{a^{-3}} \rangle_o} 
+  \frac{\langle \xi^\sigma_o | \hat{a} |\xi^\nu_o \rangle}{\langle \hat{a} \rangle_o}\right],  \quad  
\end{eqnarray}
and $A_{\mu\nu}^{(l)}$ and $B_{\mu\nu}^{(l)}$ are defined  in Eq.~(\ref{epsilon-BO}).
In the presence of   $L_1^{(l)}, L_2^{(l)}\neq0$  in 
Eq.~(\ref{speed-photon}), it is seen  that  EM modes propagate with the speed
higher than the speed of  light ($c=1$) on the  background (\ref{dressed-metric-BO}). 
By expanding Eq.~(\ref{dispersion-BO}), we get
\begin{eqnarray}
E^2\ \approx  \ \big(1+2\lambda L_1^{(l)}\big)p^2+ 2\lambda L_2^{(l)}p^4 + \mathcal{O}(p^5, \lambda^2). \quad 
\label{dispersion-BO2}
\end{eqnarray} 
Thus,  the  dispersion relation on the effective background (\ref{dressed-metric-BO})
deviates from the standard dispersion relation $E^2=p^2$. 
This deviation is held even
 in the low-energy limit  (with $k\ll 1$) since $L_1^{(l)}\neq0$
 or when the  EM field mode state (denoted by $l$) is only a vacuum state in the Fock space.
These analyzes   indicate that 
the local Lorentz symmetry is violated on the effective geometry (\ref{dressed-metric-BO}). 
The standard dispersion relation is recovered only
when the quantum gravity induced  parameters $L_1^{(l)}, L_2^{(l)}$ vanish.

\section{Particle production on emerging   background}
\label{Particle-Production}

In this section, we are interested in studying  how those quantum gravity effects, in the presence of the field's backreaction [leading to an emergent rainbow geometry (\ref{dressed-metric-BO})], can give rise to the creation of the quantum particles from the Planck regime\footnote{For the issue of particle production in quantum cosmology, see for example Ref.~\cite{Us:2015}}.

The Hamiltonian of the EM field for a background metric (\ref{dressed-metric-BO}) can be written as
\begin{eqnarray}
i\hbar\partial_{T}\psi = \frac{\tilde{N}_{T}}{2\tilde{a}^3} 
\Big[\big(\hat{P}_{\mathbf{k}}^{r}\big)^2 + k^2\tilde{a}^4\big(\hat{Q}_{\mathbf{k}}^{r}\big)^2\Big] \psi\ .
\label{Hamiltonian-SF-bquantum-mod}
\end{eqnarray}
When the background metric has no dependence on the wave number $k$ (i.e., for $\beta=\sigma=0$), then $\tilde{N}_T=\bar{N}_T$, $\tilde{a}=\bar{a}$, and Eq.~(\ref{Hamiltonian-SF-bquantum-mod}) reduces to the Hamiltonian equation (\ref{evol-eq1}). For nonvanishing backreaction effects, $\beta(k)\neq 0$ and $\sigma(k)\neq 0$, the corresponding (classical) equation of motion (\ref{motion1}) for Eq.~(\ref{Hamiltonian-SF-bquantum-mod}) is modified as
\begin{eqnarray}
\ddot{Q}_{\mathbf{k}}^{r} + \frac{\dot{\tilde{m}}}{\tilde{m}}\dot{Q}_{\mathbf{k}}^{r}+ \tilde{\omega}_{k}^2(T)Q_{\mathbf{k}}^{r}=0\ ,
\label{motion1-mod}
\end{eqnarray}
in which a dot,  here, denotes  a  derivative with respect to the internal time $T$. Now, the frequency of each mode $\tilde{\omega}_{k}(T)$ is corrected by the $k$-dependent term $f(k)$, as given in Eq.~\eqref{dispersion-BO}, and  $\tilde{m}$ is given by
\begin{eqnarray}
\tilde{m}\ =\ \frac{\tilde{a}^3}{\tilde{N}_T}\ =\ \frac{\bar{a}^3}{\bar{N}_T}\frac{(1+ \lambda\sigma(k))^3}{(1+ \lambda \beta(k))}\ .
\end{eqnarray}
This equation indicates that  $\dot{\tilde{m}}/\tilde{m}$ in Eq.~(\ref{motion1-mod}) cannot vanish anymore, because $\tilde{N}_T/\tilde{a}^3\neq 1$. 
To study the particle creation mechanism using the approach presented in Ref.~\cite{Berger:1975}, it is convenient that the equation of motion (\ref{motion1-mod}) takes the form of Eq.~(\ref{motion2}).
To do so, we cannot use the usual harmonic time gauge technique as we had at the end of Sec.~\ref{sec-BianchiI-class}, since $\dot{\tilde{m}}\neq 0$ on the emerging effective metric (\ref{dressed-metric-BO}).
However, we can consider  any new time coordinate $\tilde{\tau}$ (with a lapse function $N_{\tilde{\tau}}$) such that
$N_{\tilde{\tau}}d\tilde{\tau}=\tilde{N}_TdT$. Then, evolution in terms of the new time parameter $\tilde{\tau}$  is given by the  $\tilde{\tau}$-dependent Schr\"odinger equation:
\begin{eqnarray}
i\hbar\partial_{\tilde{\tau}}\psi = \frac{\tilde{N}_{\tilde{\tau}}}{2\tilde{a}^3} 
\Big[\big(\hat{P}_{\mathbf{k}}^{r}\big)^2 + k^2\tilde{a}^4\big(\hat{Q}_{\mathbf{k}}^{r}\big)^2\Big] \psi\ .
\label{Hamiltonian-SF-bquantum-mod-newtime}
\end{eqnarray}
Without loss of generality, we shall consider the lapse function $N_{\tilde{\tau}}$ such that $\tilde{\tau}$ is  a harmonic time, i.e., $N_{\tilde{\tau}}=\tilde{a}^3$. In this time gauge, we have that 
\begin{eqnarray}
d\tilde{\tau}\ =\ \frac{\tilde{N}_{T}}{\tilde{a}^3} dT\ =\  \frac{\bar{N}_{T}}{\bar{a}^3} \frac{(1+\lambda \beta(k))}{(1+ \lambda \sigma(k))^3} dT\ .
\end{eqnarray}
In terms of this new (harmonic) time coordinate $\tilde{\tau}$, the classical equation of motion
corresponding to the Schr\"odinger equation (\ref{Hamiltonian-SF-bquantum-mod-newtime}) becomes
\begin{eqnarray}
Q_{\mathbf{k}}^{r^{\prime\prime}} +  \tilde{\omega}_{k}^2(\tilde{\tau})Q_{\mathbf{k}}^{r}=0\ ,
\label{motion1-mod-harmonic}
\end{eqnarray}
in which a  prime  now denotes a derivative with respect to the harmonic time $\tilde{\tau}$ and $\tilde{\omega}_{k}$ is  given by 
\begin{eqnarray}
\tilde{\omega}_{k}^2(T) 
\ =\   k^2 \bar{a}^4\big[1 + \lambda \sigma(k)\big]^4 \  =\   k^2\tilde{a}^4   \ .
\label{omega-BO}
\end{eqnarray}

There exists a complete set of solutions to Eq.~(\ref{Hamiltonian-SF-bquantum-mod-newtime}) characterized by the quantum number $n$ as \cite{Perelmov:1969, Berger:1975}
\begin{eqnarray}
\chi_n(Q, \tilde{\tau}) &=& \frac{(v^{r\ast}_k)^\frac{n}{2} \mathrm{H}_n\big(\frac{Q_\mathbf{k}^r}{\sqrt{2}|v_k^r|}\big)
\exp\big[iv_k^{r^\prime} \frac{(Q_\mathbf{k}^r)^2}{2v_k^r}\big]}{
2^{\frac{n}{2}}(2\pi)^{\frac{1}{4}}(n!)^{\frac{1}{2}}  (v^{r}_k)^{\frac{n+1}{2}}}~,  \quad \quad
\label{number-state}
\end{eqnarray}
where $v_k^r(\tilde{\tau})$ is a solution of Eq.~(\ref{motion1-mod-harmonic})  and $\mathrm{H}_n$ is the Hermite polynomial of order $n$. 
The states (\ref{number-state}) can be generated by defining creation and annihilation operators $\hat{A}_\mathbf{k}^\dagger$ and $\hat{A}_\mathbf{k}$ \cite{Salusti:1970},
\begin{eqnarray}
\hat{A}_\mathbf{k}\ = \ -i v_k^{r^\prime}(\tilde{\tau}) \hat{Q}_\mathbf{k}^r + v_k^r(\tilde{\tau})(\partial/\partial Q_\mathbf{k}^r) ,
\label{annihilation}
\end{eqnarray}
where  $\hat{A}_\mathbf{k}^\dagger$ is its Hermitian conjugate of (\ref{annihilation}).
These operators have the properties 
\begin{eqnarray}
\chi_n=(n!)^{-1/2}(\hat{A}_\mathbf{k}^\dagger)^n\chi_0, \quad \quad  \quad \hat{A}_\mathbf{k}\chi_0=0 \ ,
\end{eqnarray}
where   $[\hat{A}_\mathbf{k}, \hat{A}_\mathbf{k}^\dagger]=1$.

Let us assume (following Zel'dovich \cite{Zeldovich:1970}) that there is a regime $\tilde{\tau}\leq \tilde{\tau}_i$ such that the vacuum state satisfies an adiabatic condition. Then, we can    construct the usual harmonic oscillator states and choose $|0\rangle$ as the vacuum wave function defined by
\begin{eqnarray}
\frac{1}{2}\tilde{\omega}_k(\tilde{\tau}_i)~ |0\rangle\ 
=\ \hat{H}_{\tilde{\tau},\mathbf{k}}^{(r)}(\tilde{\tau}_i)~|0\rangle\ ,
\end{eqnarray}
where $\hat{H}_{\tilde{\tau}, \mathbf{k}}^{(r)}(\tilde{\tau}_i)$ is the Hamiltonian operator given by
\begin{eqnarray}
\hat{H}_{\tilde{\tau}, \mathbf{k}}^{(r)} = -\frac{\hbar^2}{2}\frac{\partial^2}{\partial (Q_{\mathbf{k}}^{r})^2} +  \frac{1}{2} \tilde{\omega}_k^2\big(Q_{\mathbf{k}}^{r}\big)^2\ ,
\label{Hamiltonian-particle}
\end{eqnarray}
evaluated at the initial time\footnote{This can be considered  as the initial quantum bounce
at $\tilde{\tau}_i=\tilde{\tau}_{\rm B}$.} 
$\tilde{\tau}=\tilde{\tau}_i$, and 
$|0\rangle$ is the harmonic-oscillator ground state for the frequency 
$\tilde{\omega}_k(\tilde{\tau}_i)$.
In a Wentzel-Kramers-Brillouin (WKB) approximation, we expand the classical test field solution $v_k^r$ as
\begin{eqnarray}
v_{k}^r(\tilde{\tau})= \frac{1}{\sqrt{2\tilde{\omega}_k}}\exp\Big[i\int d\tilde{\tau} \tilde{\omega}_k(\tilde{\tau})\Big].
\label{WKB1}
\end{eqnarray}
If the Hamiltonian (\ref{Hamiltonian-particle}) possesses an adiabatic regime, in which 
\begin{eqnarray}
\tilde{\omega}_k^\prime\ =\ d\tilde{\omega}_k/d\tilde{\tau}\ll \tilde{\omega}_k^2~, 
\label{adiabatic-condtion}
\end{eqnarray}
then $v_k^{r^\prime}=dv_k^r/d\tilde{\tau}=i\tilde{\omega}_k v_k^r$.
In the adiabatic regime where inequality  (\ref{adiabatic-condtion}) holds, 
the annihilation operator (\ref{annihilation}) (and its Hermitian conjugate $\hat{A}_\mathbf{k}^\dagger$) reduces to the annihilation (and creation) operator  of the usual harmonic oscillator for the fixed frequency $\tilde{\omega}_k$:
\begin{eqnarray}
\hat{A}_\mathbf{k}\ &=& \ v_k^{r}(\tilde{\tau})\Big[\tilde{\omega}_k \hat{Q}_\mathbf{k}^r + (\partial/\partial Q_\mathbf{k}^r)\Big] ,  \\
\hat{A}_\mathbf{k}^\dagger\ &=& \ v_k^{r\ast}(\tilde{\tau})\Big[-\tilde{\omega}_k \hat{Q}_\mathbf{k}^r + (\partial/\partial Q_\mathbf{k}^r)\Big] . \quad
\label{annihilation2}
\end{eqnarray}
Then, we can define a (usual harmonic-oscillator) number operator as
\begin{eqnarray}
\hat{N}_k^r := \hat{A}_\mathbf{k}^\dagger\hat{A}_\mathbf{k} =  |v_k^{r}|^2\Big[ \partial^2/\partial (Q_\mathbf{k}^{r})^2 - \tilde{\omega}_k^2 (\hat{Q}_\mathbf{k}^r)^2\Big] , \quad
\label{annihilation3}
\end{eqnarray}
so that the wave functions (\ref{number-state}) are  its eigenfunctions   at all times:
\begin{eqnarray}
\hat{N}_k^r \chi_n(Q, \tilde{\tau}) = n  \chi_n(Q, \tilde{\tau})\ .
\end{eqnarray}
This fact will lead to the interpretation of the particle number in the following.

Since $\{\chi_n\}$ form a complete orthonormal set for all $\tilde{\tau}$, we can expand  the vacuum state $|0\rangle$ in the eigenstates $\chi_n$ evaluated at 
$\tilde{\tau}_i$:
\begin{eqnarray}
|0\rangle\ =\ \sum_n b_n\chi_n(Q, \tilde{\tau}_i)\ .
\end{eqnarray}
Now, we are able to calculate the expectation value of the number operator $\hat{N}_k^r$ with respect to the vacuum $|0\rangle$ as
\begin{eqnarray}
\langle 0 | \hat{N}_k^r | 0 \rangle \ =\ \frac{1}{2}\Big(\tilde{\omega}_i|v_{k}^r|^2 + \frac{|v_{k}^{r^\prime}|^2}{\tilde{\omega}_i}-1\Big),
\label{Particle-Number}
\end{eqnarray}
where $\tilde{\omega}_i=\tilde{\omega}(\tilde{\tau}_i)$.

If the condition (\ref{adiabatic-condtion}) is valid for all $\tilde{\tau}\geq \tilde{\tau}_{i}$, so that the system is always adiabatic, the number of particle production (\ref{Particle-Number}) for the general solution (\ref{WKB1}) can be expanded as
\begin{eqnarray}
\langle 0 | \hat{N}_k^r | 0 \rangle \ \approx \ \frac{1}{4}\Big(\frac{\tilde{\omega}_i}{\tilde{\omega}_k} + \frac{\tilde{\omega}_k}{\tilde{\omega}_i} -2\Big) + {\cal O}(\tilde{\omega}_k^\prime/\tilde{\omega}_k^2). \quad
\label{Particle-Number-WKB}
\end{eqnarray}
In the case of test-field approximation (i.e., in the absence of backreaction effects), as long as the adiabaticity condition on the background spacetime holds, the number of particle creation (\ref{Particle-Number})  reduces to\footnote{cf. Ref.~\cite{Berger:1975} for the case in which  the background is purely classical.}
\begin{eqnarray}
\langle 0 | \hat{N}_k^r | 0 \rangle\ =\  0 + {\cal O}(\tilde{\omega}_k^\prime/\tilde{\omega}_k^2)\ .
\end{eqnarray}
Thus, when the system is always adiabatic, (almost) no particles are created \cite{Berger:1975}.
However, in the presence of backreaction effects (following the Born-Oppenheimer approximation),
the form of the Schr\"odinger equation (\ref{Hamiltonian-SF-bquantum-mod-newtime}) is different from the classical one (\ref{Hamiltonian-SF-bquantum}) so that we expect a modified number of particle production [up to the leading-order terms in $\sigma(k)$] as
\begin{eqnarray}
\langle 0 | \hat{N}_k^r | 0 \rangle \ \approx \ \lambda^2 \sigma^2 + {\cal O}(\lambda^3) + {\cal O}(\tilde{\omega}_k^\prime/\tilde{\omega}_k^2).
\label{particle-1polar}
\end{eqnarray}
Note that, for each mode, the number of created particles must be twice  the result given by Eq.~(\ref{particle-1polar}) because of two polarization scalar modes. So, each mode contains 
\begin{eqnarray}
\langle 0 | \hat{N}_k | 0 \rangle\  =\  \sum_r^2 \langle 0 | \hat{N}_k^r | 0 \rangle\  \approx \ 
2\lambda^2 \sigma^2  \quad \quad 
\label{particle-2polar}
\end{eqnarray}
amount of particles in the adiabatic limit.
Therefore,   backreaction of the EM field  on the quantum geometry can give rise to the particle production rate (\ref{particle-2polar}).
The number of created  particles depends on the energy of the field modes. So, modes with higher energies probe higher amount of particle creation. Notice that  we have considered only the first-order perturbations in $\lambda$ in the mode eigenenergy, but considering higher than quadratic-order terms of $\lambda$ in perturbations might lead to the higher creation amount.

\section{Conclusion}
\label{Conclusion}

In this paper, we considered an  EM  field  propagating on a classical flat FLRW spacetime coupled to an irrotational dust field $T$ as a background matter source. 
We  showed  that the Hamiltonian of the EM field can be written as the  Hamiltonian of decoupled harmonic oscillators, each corresponding to a single mode and polarization of the field. 
Using the well-known procedure of loop quantum cosmology coupled to dust \cite{Pawlowski:2012}, 
the background geometry  was  quantized, in which  $T$ played  the role of time in quantum theory. 
In quantum theory, the total Hamiltonian constraint was solved in order to get a $T$ evolution  for the total wave function of the system. To make a comparison between the evolution equations of the field wave function on the classical background with that given on the quantum background geometry,  we employed two different techniques, namely,  the test-field and the Born-Oppenheimer  approximations.

By applying a test-field approximation into the total wave function of the system, we  disregarded  the backreaction between the field and the geometry. We obtained an effective (classical) background on which quantum modes of the EM field propagate. This emerging effective background has the same form of the original FLRW spacetime, but with the metric components depending on the quantum  fluctuation of the original  background geometry.
Nevertheless, by taking  the backreaction effects into consideration, through  a Born-Oppenheimer approximation, we  found an effective geometry emerging,  the  components of which  depend on the both quantum geometry fluctuations and the modes of the field. More precisely,  EM fields with different energies probe different backgrounds provided, leading to a rainbow metric forming. This result gives rise to  violation of the local Lorentz symmetry on the dressed geometry.

Finally, we presented a mechanism for particle production in an adiabatic regime on the resulting effective backgrounds. Our computation showed that the amount of particle creation depends on the  modifications provided by backreaction effects through Born-Oppenheimer approximation. In particular, when backreaction was  discarded, no amount of particles could  be detected. However, when the backreaction was taken into account, EM field modes could probe particle creations. Moreover, the number of particle production depended  on the mode of the field; that is, modes with higher energy detected a higher amount of  particle production and vice versa.

\section*{Acknowledgments}

The authors would like to thank N. Ahmadi  for the useful comments  
on quantum theory of EM field in a classical background.
M.~N-Z. and A.~P. thank the research council of the University of Tehran for financial support.
Y.~T. acknowledges  Bonyad-e-Melli Nokhbegan of Iran (INEF) for  financial support, and the  Brazilian agencies FAPES and CAPES for partial financial support. He  thanks   Chun-Yen Lin, T. Pawlowski, and S. Sheikh-Jabbari for very useful discussions on this  work. He is also  grateful  for  the warm hospitality of the University of Warsaw where part of this work was completed.  
This work was supported by the grant of Polish Narodowe Centrum Nauki, Grant No. 2011/02/A/ST2/00300.

\appendix

\section{Connection coefficients and Ricci tensor in the FLRW spacetime}
\label{cf2}

For  the metric \eqref{metric-class},
the nonzero connection coefficients and Ricci tensor in the rectangular coordinates are given by the following expressions:
\begin{eqnarray}
&& {\Gamma^0}_{00}=\frac{\dot{N}}{N}, \quad \quad {\Gamma^0}_{xx}={\Gamma^0}_{yy}={\Gamma^0}_{zz} =\frac{a\dot{a}}{N^2}\ , \quad \quad 
\nonumber \\ && {\Gamma^x}_{0x}={\Gamma^y}_{0y}={\Gamma^z}_{0z}=\frac{\dot{a}}{a} \ ,
\end{eqnarray}
and
\begin{eqnarray}
&& \mathcal{R}_{00}=\frac{3\ddot{a}N-\dot{N}\dot{a}}{aN}\ , \nonumber \\  
&& \mathcal{R}_{xx}=\mathcal{R}_{yy}=\mathcal{R}_{zz}=-\frac{a\ddot{a}N-a\dot{N}\dot{a}+2N{\dot{a}}^2}{N^3} \ \cdot \quad \quad 
\end{eqnarray}
These  connection coefficients  indicate   that  there are no terms involving first-order spatial derivatives of the components of EM potential in the wave equation \eqref{Maxwell2}.

\section{Hamiltonian of the EM field in the FLRW spacetime}
\label{Hamiltonian}

We can rewrite the Lagrangian density (\ref{Lagrangian})  as
\begin{eqnarray}
{\cal L}_{\rm EM}  =  -\frac{1}{2}\sqrt{-g}\Big[\sum_i^3 F_{0i}F^{0i} + \sum_{i<j}^3 F_{ij}F^{ij}\Big].
\label{Lagrangian1}
\end{eqnarray}
By fixing the gauge condition $A^0=0$, for which $F_{0i}=\partial_0A_i$ and $F^{0i}=\partial^0A^i$, Eq.~(\ref{Lagrangian1}) reduces to
\begin{eqnarray}
{\cal L}_{\rm EM} &=& -\frac{1}{2}\sqrt{-g}\Big[\sum_i \sum_j(\partial_0A_i) g^{00}g^{ij} (\partial_0A_j) \nonumber \\
&& \quad \quad \quad \quad \quad \quad\quad\quad  + \sum_{i<j}  F_{ij}  F^{ij}\Big]. 
\label{Lagrangian1a}
\end{eqnarray}
The conjugate momentum $\pi^i$ for the vector fields $A_i$ can be defined as
\begin{eqnarray}
\pi^0 &=&  \frac{\delta\mathcal{L}_{\rm EM}}{\delta(\partial_0A_0)}=0\ , \\
\pi^i &=&  \frac{\delta\mathcal{L}_{\rm EM}}{\delta(\partial_0A_i)} = -\sqrt{-g} F^{0i}\ =\  E^i\  .
\label{E-i}
\end{eqnarray}
Notice that  $\pi^i$ is conjugate to $A_i=g_{ij}A^j=a^2A^i$.
Then, the usual commutation relations read
\begin{eqnarray}
[A_i(x_0, \mathbf{x}), \pi^j(x_0, \mathbf{x}^\prime)]= i {\delta^i}_{j}\ .
\end{eqnarray}
By using $\mathcal{H}_{\rm EM}=\pi^i(\partial_0A_i)-\mathcal{L}_{\rm EM}$, 
the Hamiltonian density is  written as
\begin{eqnarray}
\mathcal{H}_{\rm EM} =  \frac{1}{2}\sqrt{-g}\Big[\sum_{i} \frac{g_{00}}{g}  
(\pi_i \pi^i) + \sum_{i<j}  F_{ij} F^{ij}\Big].
\label{Hamiltonan-den1a}
\end{eqnarray}
For the  flat FLRW background (\ref{metric-class}), the Hamiltonian density of EM field (\ref{Hamiltonan-den1a}) reduces to
\begin{eqnarray}
\mathcal{H}_{\rm EM}\ =\ \frac{N_{x_0}}{2a^3}\Big[\sum_{i}\pi_i \pi^i +
a^6\sum_{i<j}F_{ij}F^{ij}\Big] .
\label{Hamiltonan-den1}
\end{eqnarray}

To calculate the last  term in the    Hamiltonian  density (\ref{Hamiltonan-den1}), we write
\begin{eqnarray}
\sum_{i<j}F_{ij}F^{ij} &=& F_{12}F^{12} + F_{13}F^{13} + F_{23}F^{23} \nonumber \\
 &=&  \sum_{i, j}^3\Big[(\partial_iA_j)(\partial^iA^j) - (\partial_iA_j)(\partial^jA^i)\Big]. \quad \quad
\label{sec-term}
\end{eqnarray}
We define the covariant derivative in three-dimensional space  in the same way we define the four-dimensional covariant derivative; then, $A_{a|b} $ shows the three-dimensional  covariant  derivative in terms of partial derivatives and connections of the given 3-space ($\Sigma, \gamma $),
\begin{eqnarray}
\nabla_j A^i\ :=\ A^i_{~|j}\ =\ \partial_j A^i+r^i_{~jk} A^k\ ,
\end{eqnarray}
where $r^i_{~jk}$ are connections of the 3-space, 
and the 3-dimensional divergence of $\textbf{A}$ reads
\begin{eqnarray}
\nabla \cdot \mathbf{A}\ =\  A^i_{~|i}=\dfrac{1}{\sqrt{\gamma}} \dfrac{\partial}{\partial_i} (\sqrt{\gamma} A^i)\ .
\end{eqnarray}
In terms of the three-dimensional  covariant  derivative, Eq.~(\ref{sec-term}) takes the following form:
\begin{eqnarray}
F_{ij}F^{ij} &=& 2 \Big[ (\partial_iA_j)(\partial^iA^j) - (\partial_iA_j)(\partial^jA^i)\Big] \nonumber \\
&=& 2 \Big[ \nabla_iA_j \nabla^iA^j - \nabla_iA_j \nabla^jA^i \Big]\ .
\label{detail-A}
\end{eqnarray}
The second term in the equation above can be written as a total divergence:
\begin{eqnarray}
(\nabla_iA_j)(\nabla^jA^i) \ =\ \nabla_i(A_j\nabla^jA^i) - A_j(\nabla_i\nabla^jA^i).
\nonumber
\end{eqnarray}
In the summation, the last term in the equation above reads 
$ A_j(\nabla^j\nabla_iA^i)=A_j\nabla^j (\nabla \cdot \mathbf{A})$.
Then, in the {\em radiation gauge}, $\nabla \cdot \mathbf{A}=0$, the last term vanishes. Thus, the total divergence is equal to the first term only, which  vanishes by integration with respect to the spatial volume when computing the total Hamiltonian $H_{\rm EM}$, so that  only the first term in 
Eq.~(\ref{sec-term}) remains nonzero. 
On the other hand,  for the first term, we can also write
\begin{eqnarray}
(\nabla_iA_j)(\nabla^iA^j)  \ =\ \nabla_i(A_j\nabla^iA^j)  -  A_j\nabla_i\nabla^iA^j.
\nonumber
\end{eqnarray}
Using the divergence theorem in three-dimensional  curved space, we have that
\begin{eqnarray}
\int_\Sigma \sqrt{\gamma} ~ dx^3 A^i_{~|i} =  \oint_{\partial\Sigma} A^i d\Sigma_i\ , \nonumber
\end{eqnarray}
where $\gamma=a^6$ herein our  background spacetime.  Then, we obtain
\begin{eqnarray}
\int \sqrt{\gamma} dx^3  (\nabla_iA_j)(\nabla^iA^j) = - \int \sqrt{\gamma} dx^3 A_j\nabla^2A^j\  .
\nonumber
\end{eqnarray}

Using the ingredients provided in the paragraph above, by integrating over the Hamiltonian density  (\ref{Hamiltonan-den1}), 
we obtain the total Hamiltonian (\ref{Hamilton-app1}) for the EM field.
The total Hamiltonian (\ref{Hamilton-app1}) can be split  through positive and negative sectors on the lattice ${\cal L}$ as
\begin{eqnarray}
H_{\rm EM} &=& \sum_{\mathbf{k}\in {\cal L}_+} \sum_r  H_{\mathbf{k}}^{(r)+} + \sum_{\mathbf{k}\in {\cal L}_-} \sum_r  H_{\mathbf{k}}^{(r)-} \nonumber \\ 
&=&  \frac{N_{x_0}}{2a^3} \sum_r  \Big[\sum_{\mathbf{k}\in {\cal L}_+} \Big(\big(\pi_{\mathbf{k}}^{r}\big)^\ast\pi_{\mathbf{k}}^{r} + 
 k^2a^4\big(\mathrm{A}_{\mathbf{k}}^{r}\big)^\ast \mathrm{A}_{\mathbf{k}}^{r}\Big)  
 \nonumber \\
 && \quad \quad   + \sum_{\mathbf{k}\in {\cal L}_-} \Big(\big(\pi_{\mathbf{k}}^{r}\big)^\ast\pi_{\mathbf{k}}^{r} + 
 k^2a^4\big(\mathrm{A}_{\mathbf{k}}^{r}\big)^\ast \mathrm{A}_{\mathbf{k}}^{r}\Big) \Big] \notag \\
&=&   \frac{N_{x_0}}{2a^3} \sum_r  \sum_{\mathbf{k}\in {\cal L}_+} \Big[\big(\pi_{\mathbf{k}}^{r}\big)^\ast\pi_{\mathbf{k}}^{r} +\big(\pi_{-\mathbf{k}}^{r}\big)^\ast\pi_{-\mathbf{k}}^{r} \nonumber \\
 && \quad \quad  +  k^2a^4\Big(\big(\mathrm{A}_{\mathbf{k}}^{r}\big)^\ast \mathrm{A}_{\mathbf{k}}^{r} + \big(\mathrm{A}_{-\mathbf{k}}^{r}\big)^\ast \mathrm{A}_{-\mathbf{k}}^{r}\Big)  \Big]. \quad \quad
\label{Hamilton-app-b1}
\end{eqnarray}
In the last step above,  by a simple changing of  index $\mathbf{k}\rightarrow-\mathbf{k}$,
we  have converted   the sum over the negative index to that over  the positive index.
Substituting now $\mathrm{A}_{\mathbf{k}}^{r}$ and 
$\pi_{\mathbf{k}}^{r}$ from
 Eqs.~(\ref{app-phi-pi-1a}) and (\ref{app-phi-pi-1b}) into Eq.~(\ref{Hamilton-app-b1}),  we obtain
\begin{eqnarray}
H_{\rm EM}  &=& \frac{N_{x_0}}{4a^3}\sum_{\mathbf{k}\in {\cal L}_+}
 \sum_r 
\sum_{n=1}^2 \Big[\big(\pi_{\mathbf{k}}^{r(n)}\big)^2 
+ \big(\pi_{-\mathbf{k}}^{r(n)}\big)^2  \nonumber \\
 && \quad \quad \quad \quad  +  k^2a^4\Big(\big(\mathrm{A}_{\mathbf{k}}^{r(n)}\big)^2 
+ \big(\mathrm{A}_{-\mathbf{k}}^{r(n)}\big)^2\Big)  \Big] \notag \\
&=&  \frac{N_{x_0}}{2a^3}\sum_{\mathbf{k}\in {\cal L}_+}\sum_r 
\sum_{n=1}^2 \Big[\big(\pi_{\mathbf{k}}^{r(n)}\big)^2 
 +  k^2a^4\big(\mathrm{A}_{\mathbf{k}}^{r(n)}\big)^2  \Big] . \nonumber \\
\label{Hamilton-app-b2}
\end{eqnarray}
This equation represents  a  Hamiltonian for a  collection  of independent harmonic oscillators. 
We can further rewrite the last relation in Eq.~(\ref{Hamilton-app-b2})  as  sum of modes of two harmonic oscillators:
\begin{eqnarray}
H_{\rm EM} &=&  \frac{N_{x_0}}{2a^3}\sum_{\mathbf{k}\in {\cal L}_+}\sum_r  \Big\{\Big[\big(\pi_{\mathbf{k}}^{r(1)}\big)^2 
 +  k^2a^4\big(\mathrm{A}_{\mathbf{k}}^{r(1)}\big))^2  \Big]  \nonumber \\
 && \quad \quad \quad \quad  + \Big[\big(\pi_{\mathbf{k}}^{r(2)}\big)^2 
 +  k^2a^4\big(\mathrm{A}_{\mathbf{k}}^{r(2)}\big))^2  \Big] \Big\} \notag \\
 &=&  \frac{N_{x_0}}{2a^3}\sum_{\mathbf{k}\in {\cal L}_+}\sum_r \Big[\big(\pi_{\mathbf{k}}^{r(1)}\big)^2 
 +  k^2a^4\big(\mathrm{A}_{\mathbf{k}}^{r(1)}\big)^2 \Big]  \nonumber \\
   &&   +\frac{N_{x_0}}{2a^3}\sum_{\mathbf{k}\in {\cal L}_-}\sum_r  \Big[\big(\pi_{-\mathbf{k}}^{r(2)}\big)^2 
 +  k^2a^4\big(\mathrm{A}_{-\mathbf{k}}^{r(2)}\big)^2 \Big] .\quad\quad  \quad
\label{Hamilton-app-b3}
\end{eqnarray}
Now, by defining  the new real variables (\ref{def-q}) and (\ref{def-p}) and substituting in  Eq.~(\ref{Hamilton-app-b3}), 
the total Hamiltonian (\ref{Hamiltonian-SF1}) is obtained.


\end{document}